\documentclass[12pt]{iopart}
\usepackage{graphicx}% Include figure files
\usepackage{dcolumn}% Align table columns on decimal point
\usepackage{bm}% bold math
\usepackage{subfigure}
\usepackage{marvosym}
\usepackage{textcomp}
\usepackage{threeparttable}
\usepackage{booktabs}
\usepackage{slashed}
\usepackage{citesort}
\usepackage{color}

\begin{document}

\title{$J/\psi$ Production by Magnetic Excitation of $\eta_c$}
\author{Di-Lun Yang and Berndt M\"uller}
\address{Department of Physics, Duke University, Durham, NC 27708, USA}
\date{\today}

\begin{abstract}
We compute the probability of $J/\psi$ production from the interaction between $\eta_c$ and the strong magnetic field generated in relativistic heavy ion collisions. The computation is first carried out in the heavy quark effective model, in which the M1 radiative transition is considered. Then we investigate the transition in the framework of non-relativistic heavy hadron chiral perturbation theory and show that the polarization of $J/\psi$ produced by this process is parallel to the direction of magnetic field and thus perpendicular to the reaction plane. The transition probability obtained in both approaches is of order $2\times 10^{-4}$.
\end{abstract}

\section{Introduction}
The production of heavy quarkonia in relativistic heavy ion collision has been widely studied in consideration of their possible role as probes of the quark-gluon plasma (QGP) and the QCD deconfinement transition. $J/\psi$ is the most promising candidate among the charmonium probes because of its relatively large production cross section and its large branching ratios for leptonic decays. Various competing mechanisms contribute to the production and the suppression of $J/\psi$ in nucleus-nucleus collisions. The so-called ``anomalous'' suppression of $J/\psi$ was first observed in experiments at the CERN-Super Proton Synchrotron (SPS) by the NA38 collaboration \cite{Baglin:1990iv}, then confirmed by the NA50 and NA60 collaborations \cite{Abreu1997327,Arnaldi:2007aa}, and later observed in experiments at the Relativistic Heavy Ion Collider (RHIC) by the PHENIX collaboration \cite{PhysRevLett.96.012304}. 

Theoretically, charmonium states are successfully described as nonrelativistic bound states of a $c\bar{c}$ pair with a binding potential incorporating one-gluon exchange and a linear confining potential. Matsui and Satz \cite{Matsui:1986dk} proposed that the color screening of the quark-antiquark potential in a quark-gluon plasma will hinder the formation of charmonia. Their prediction of the suppression of charmonium production preceded the experimental observations; however, the measurements did not confirm the anticipated suppression pattern. In order to understand the experimental data in detail, various other mechanisms for charmonium suppression have been investigated, so far without a generally accepted, comprehensive solution. In parallel, the color screening effect has been studied in several different approaches. Recent results on color screening based on calculations in lattice QCD can be found in \cite{Mocsy:2007jz}.

Besides the screening effect, the suppression may be caused by the kinetic dissociation of $c\bar{c}$ pairs. Due to finite formation time of charmonia, while a $c\bar{c}$ pair is produced in the collision, the pre-charmonium state may interact with the hard gluons when traversing the medium and dissociate \cite{Wittmann:1992qx,Grandchamp:2003uw}. In the hadronic phase, the $c\bar{c}$ pair may combine with light quarks to form a pair of heavy mesons \cite{Lin:2003jy}. Moreover, the production of $J/\psi$ can potentially be influenced in the late-phase hadronic environment by scattering with other hadrons and dissociation into a $D\bar{D}$ pair \cite{Matinyan:1998cb}.  These effects will also contribute to the suppression.

In relativistic heavy ion collisions, $c\bar{c}$ pairs are created in the hard scattering of two gluons, and different charmonium states may be formed when the resonant energy is reached and the correct quantum numbers are established. For example, according to Yang's theorem \cite{Yang:1950rg}, $\eta_c$ and $J/\psi$ can be produced through $gg\rightarrow \eta_c$ and $gg\rightarrow J/\psi g$ respectively. Nevertheless, not all charmonia are directly produced in the collision. As the energy of collision increases, the number of $c\bar{c}$ pairs created by hard scattering grows. The recombination of these charm and anti-charm quarks from different binary nucleon collisions may cause a considerable enhancement of the charmonium production. Based on the hydrodynamic simulation of the central collision under RHIC condition, between 30$\%$ and 50$\%$ of the $J/\psi$ yield may come from the recombinant production \cite{Zhao:2008pp,Young:2009tj}. For instance, through the measurement of the $e^+e^-$ decay mode, the decay chain of $\chi_c(\psi')\rightarrow J/\psi\gamma\rightarrow e^+e^-\gamma$ can be reconstructed in p+p collision \cite{Oda:2008zz}. Furthermore, the suppression of excited charmonia may also result in the reduced production of $J/\psi$ and other low energy states in the charmonium spectrum due to their greater fragility in the medium. 

Motivated by the $J/\psi$ production through charmonium decays, we here investigate the $J/\psi$ production through the radiative M1 transition of $\eta_c$ induced by the strong magnetic field generated in the ultra-relativistic A+A collision. The ability of such transient fields to mix pseudoscalar and vector mesons in the light quark sector was recently discussed by Asakawa {\em et al.} \cite{Asakawa:2010bu} in connection with the chiral magnetic effect \cite{Fukushima:2008xe}. The same mechanism also induces a magnetic dipole transition between the pseudoscalar and vector states of heavy quarkonia.  Here, we first compute the transition probability $\eta_c+B\rightarrow J/\psi$, where $B$ denotes the magnetic field, by considering the magnetic dipole interaction in the quark model. Then we confirm the computation by utilizing non-relativistic heavy hadron chiral perturbation theory (NRHH$\chi$PT  \cite{Hu:2005gf}) to the leading order. Finally, we estimate the transition probability in a Au+Au collision at RHIC.      

\section{Radiative M1 transition}

To analyze the transition of $J/\psi$ from the interaction between $\eta_c$ and magnetic field generated in the collision, we consider the corona scenario which applies to the hadronic matter remaining near the surface of the colliding nuclei \cite{Asakawa:2010bu}. An $\eta_c$ meson produced near the surface can still interact with the magnetic field generated during the early time when the magnetic field is very strong. We choose the $x-y$ plane as the reaction plane and assume that the beam is oriented along the $x$ direction. We also assume the nuclear charge $Ze$ to be distributed uniformly within the nuclear radius $R$. In peripheral collisions, when the impact parameter $b\approx R$, the magnetic field at $x=0$ on the surface can be approximated as
\begin{eqnarray}
{\bf B}(t)=\frac{Ze\gamma vb}{4\pi(R^2+\gamma^2v^2t^2)^{3/2}}\hat{z},
\label{eq1}
\end{eqnarray}
where $\gamma=1/\sqrt{1-v^2}$ is the Lorentz contraction factor. The details of approximation can be found in the Appendix of  \cite{Asakawa:2010bu}. Since the collision is ultrarelativistic, the two ions can be visualized as two colliding pancakes due to the Lorentz contraction in the $x$ direction. Therefore, we expect the region to be  in the vicinity of $x=0$.

We now outline the computation of transition probability in the quark model. Since $\eta_c$ and $J/\psi$ are pseudoscalar and vector mesons, respectively, conservation of parity and charge conjugation symmetry dictates that the transition will be governed by the magnetic dipole interaction. By using the time dependent perturbation theory, the transition probability amplitude can be written as,
\begin{eqnarray}
a(t)=-i\int^{t}_{-\infty}e^{iE_{fi}t'} \langle f|H_{\rm int}(t')|i \rangle dt',
\label{eq2}
\end{eqnarray}
where $E_{fi}=E_f-E_i$ is the energy difference of the initial and final states. In our case, $\langle f|=\langle J/\psi|$, $|i\rangle =|\eta_c\rangle $ and $H_{int}(t)=-{\bf \mu}_c\cdot{\bf B}(t)$, where  ${\bf \mu}_c=\frac{e_ce}{m_c}({\bf S}_c-{\bf S}_{\bar{c}})$ is the magnetic dipole moment of the $c\bar{c}$ pair. Here $e_c=2/3$ and $m_c$ denote the charge and mass of the $c$-quark, respectively, and ${\bf S}_{c(\bar{c})}$ denotes the spin of the $c$-($\bar{c}$-) quark. We neglect strong interaction corrections to the $g$-factor of $c$-quark. By taking the limit $t\to\infty$, we obtain the total transition amplitude in the quark model approach as
\begin{eqnarray}
\nonumber
a_{fi}^{\rm QM} &=& i\frac{Ze\gamma vb}{4\pi}\frac{2e}{3m_c}\langle f|S_{cz}-S_{\bar{c}z}|i\rangle\int^{\infty}_{-\infty}\frac{e^{iE_{fi}t'}}{(R^2+v^2\gamma^2t'^2)^{3/2}}dt' 
\\ 
&=&i\frac{Ze\gamma vb}{4\pi}\, \frac{2e}{3m_c}\,
\frac{2E_{fi}\, K_1\left(\frac{E_{fi}R}{v\gamma}\right)}{v^2\gamma^2R},
\label{eq3}
\end{eqnarray}
where $K_1$ is the modified Bessel function of second kind and 
\begin{equation}
\langle f |S_{cz}-S_{\bar{c}z}| i\rangle =\langle J_f,M_f |S_{cz}-S_{\bar{c}z}| J_i,M_i \rangle =1
\label{eq4}
\end{equation}
for $J_i=0, J_f=1$ and $M_i=M_f=0$. The transition probability is given by
\begin{equation}
|a_{fi}^{\rm QM}|^2 =  \frac{16\pi\alpha}{9m_c^2}\frac{Z^2\alpha\gamma^2 v^2b^2}{4\pi}
\left(\frac{2E_{fi}\, K_1\left(\frac{E_{fi}R}{v\gamma}\right)}{v^2\gamma^2R}\right)^2.
\label{eq5}
\end{equation} 
The prefactor $16\pi\alpha/(9m_c^2)$ represents the strength of the coupling between the charmonium spin and the magnetic field. A similar result will be expected in the effective field theory (EFT) calculation, which we perform next, while the prefactor will be replaced by a coupling constant which encodes the non-perturbative physics of the charmonium bound state.

\section{Calculation in the effective field theory}
 
To calculate the transition probability in the formalism of EFT, we introduce an effective Lagrangian that describes the interaction between $\eta_c$, $J/\psi$ and the magnetic field. Since the hyperfine splitting of $\eta_c$ and $J/\psi$ is about 100 MeV, the transition can be described non-relativistically in the rest frame of the $J/\psi$. We thus make use of the framework of non-relativistic heavy hadron chiral perturbation theory (NRHH$\chi$PT  \cite{Hu:2005gf}). By preserving parity, charge conjugation, and chiral symmetry, the leading order interaction term can be written as,
\begin{eqnarray}
\mathcal{L}_{int}=\frac{\rho}{2}Tr[J {\bf B} \cdot{\bf \sigma}J^{\dagger}],
\label{eq6}
\end{eqnarray}
where $\rho$ is a coupling constant and $J={\bf \psi}\cdot{\bf \sigma}+\eta_c$ is the $\eta_c-J/\psi$ chiral multiplet. Due to the presence of the Pauli matrix, this term can break the heavy quark spin symmetry. Then the transition amplitude can be written as
 \begin{eqnarray}
\nonumber
T_{fi} &=& \langle J/\psi,B_{\rm ex}|\eta_c,B_{\rm ex}\rangle\\
&=&\nonumber
ig' \int d{\bf p}_f\, d{\bf p}_i\, \phi_{J/\psi}^*({\bf p}_f)\, \phi_{\eta_c}({\bf p}_i) 
\\ \nonumber & & \times
\int d^4x\, \langle J/\psi,B_{\rm ex}|\psi^{\dagger}_{\mu}(x) \eta_c(x){B}_{\mu}(x)|\eta_c,B_{\rm ex}\rangle 
\\ \nonumber
&=& ig' \int d{\bf p}_f\, d{\bf p}_i\, \phi_{J/\psi}^*({\bf p}_f)\, \phi_{\eta_c}({\bf p}_i)
\\ \nonumber & & \times
\langle J/\psi,B_{\rm ex}|\psi^{\dagger}_{\mu}(0) \eta_c(0)|\eta_c,B_{\rm ex}\rangle 
\\ \nonumber & & \times
\int d^4xe^{i(p_f-p_i)x}B_{\rm ex}(x)_{\mu}
\\ 
&=& ig' \int d{\bf p}_f\, d{\bf p}_i\, \phi_{J/\psi}^*({\bf p}_f)\, \phi_{\eta_c}({\bf p}_i)
\tilde{B}_{\rm ex}({\bf p}_f-{\bf p}_i)\cdot\epsilon^{J/\psi},\label{eq7}
\end{eqnarray}
where $d{\bf p}_f = d^3p_f/[(2\pi)^3\sqrt{2E_f}]$ is the relativistic phase space measure, and $\phi_{J/\psi}^*({\bf p}_f)$ and $\phi_{\eta_c}({\bf p}_i)$ represent wave packets of the $J/\psi$ and $\eta_c$. By taking the plane-wave approximation, $\phi_{J/\psi}^*({\bf p}_f)$ and $\phi_{\eta_c}({\bf p}_i)$ will be simply identity. The subscript $\mu$ denotes the direction of polarization, $\epsilon^{J/\psi}$ is the polarization vector of $J/\psi$, and $\tilde{B}_{\rm ex}$ is the four-dimensional Fourier transform of the external magnetic field.  Finally, $g'=2\sqrt{m_{J/\psi}m_{\eta_c}}\rho$ takes into account the normalization of the non-relativistic heavy fields. The unknown parameter $\rho$ can be extracted from analyzing the decay width of $J/\psi\rightarrow \gamma\eta_c$ in quark model \cite{Voloshin:2007dx}.  

In order account for the Lorentz contraction of the magnetic field (\ref{eq1}) in the $x$-direction, we now keep its full $x$-dependence. For simplicity, we consider the production at $y=0$ near the nuclear surface. The $z$-component of the magnetic field contributed by a single nucleus can be written as 
\begin{eqnarray}
B_z^{(\pm)}(t,x)=\frac{Ze\gamma b}{8\pi(R^2+\gamma^2(x\mp vt)^2)^{3/2}},
\label{eq8}
\end{eqnarray}
where the minus (plus) sign is for a right-(left-)moving nucleus. We firstly evaluate the contribution for the right-moving nucleus. By taking $v=1$ and choosing the light-cone coordinate, we can easily derive the magnetic field in momentum space,
\begin{eqnarray}
\tilde{B}^{(+)}_{\rm ex}({\bf p}_f-{\bf p}_i) &=& \frac{Ze\gamma vb}{8\pi}\, (2\pi)^3\delta(E_{fi}-p_{fi})
\nonumber \\ & & \times
\delta^2({\bf p}_{f\perp}-{\bf p}_{i\perp})(E_{fi}+p_{fi})
\frac{K_1\left(\frac{(E_{fi}+p_{fi})R}{2\gamma}\right)}{\gamma^2R},
\label{eq9}
\end{eqnarray}
where $p_{fi} = p_{fx}-p_{ix}$ and $\delta(E_{fi}-p_{fi})$ can be rewritten as 
\begin{eqnarray}
&& \hspace{-5mm} 
\frac{(E_i-p_{ix})^2+p_{f\perp}^2+m_f^2}{2(E_i-p_{ix})^2}\delta\left(p_{fx}-\frac{p_{f\perp}^2+m_f^2-(E_i-p_{ix})^2}{2(E_i-p_{ix})}\right).
\label{eq10}
\end{eqnarray} 
Since $m_{J/\psi}\approx m_{\eta_c}\gg p_f,p_i$ non-relativistically, we will make the following approximation to simplify the expression:
\begin{eqnarray}
\delta\left( p_{fx}\phantom{\frac{1}{2}}\hspace{-10pt} \right. 
  - \left.\frac{p_{f\perp}^2+m_f^2-(E_i-p_{ix})^2}{2(E_i-p_{ix})}\right)
&\approx & \delta\left(p_{fx}-\frac{m_f^2-m_i^2+2p_{ix}m_i}{2m_i}\right)
\nonumber \\
&\approx & \delta(p_{fx}-p_{ix}-E_{fi})
\label{eq11}
\end{eqnarray} 
and 
\begin{equation}
(E_i-p_{ix})^2+p_{f\perp}^2+m_f^2\approx m_{J/\psi}^2+m_{\eta_c}^2.
\label{eq12}
\end{equation} 
Inserting $\tilde{B}_{\rm ex}({\bf p}_f-{\bf p}_i)$ into (\ref{eq7}), we obtain the transition probability amplitude,
\begin{eqnarray}
a^{(+)}_f &\approx & ig'\frac{Ze\gamma b}{8\pi}
\int \frac{d{\bf p}_i}{\sqrt{2E_f}}\, \phi_{J/\psi}^*({\bf p}_f)\, \phi_{\eta_c}({\bf p}_i)
\left(\frac{m_{J/\psi}^2+m_{\eta_c}^2}{2m_{\eta_c}^2}\right)
\nonumber \\ & & \times
2E_{fi}\, \frac{K_1\left(\frac{E_{fi}}{\gamma}R\right)}{\gamma^2R}|_{{\bf p}_f=(p_{ix}+E_{fi},{\bf p}_{i\perp})}.
\label{eq13}
\end{eqnarray} 
Repeating the calculation for the minus sign in (\ref{eq8}),  we find $a_f^{(+)}=a_f^{(-)}$. If we only keep the leading-order term in the expansion of $E_{fi}$, the full nonrelativistic transition probability can be expressed as,
\begin{eqnarray}
\nonumber
|a_{fi}^{\rm NR}|^2 &=& \frac{g'^2}{4E_iE_f}\left(\frac{Ze\gamma b}{4\pi}\right)^2 
\left(\frac{2E_{fi}\, K_1\left(\frac{E_{fi}R}{\gamma}\right)}{\gamma^2R}\right)^2
\\
&\approx & \rho^2\frac{m_{\eta_c}}{m_{J/\psi}}\, \frac{Z^2\alpha\gamma^2 b^2}{4\pi}
\left(\frac{2E_{fi}\, K_1\left(\frac{E_{fi}}{\gamma}R\right)}{\gamma^2R}\right)^2,\label{eq14}
\end{eqnarray}
which is equal to the transition probability one finds by just using the magnetic field expression (\ref{eq1}). This correspondence reflects the fact that produced $J/\psi$ mesons are highly localized at $x=0$.

In the previous calculation, we analyzed the transition in the rest frame of $J/\psi$. Nevertheless, the $\eta_c$ directly produced in the collision may move relativistically. Under this circumstance, the NRHH$\chi$PT has to be promoted to its relativistic version\cite{Casalbuoni:1992yd} and the charmonium field must be rewritten as 
\begin{eqnarray}
J=\frac{1}{2}(1+\slashed v)[\psi_{\mu}\gamma^{\mu}-\eta_c\gamma_5]\frac{1}{2}(1-\slashed v),
\label{eq15}
\end{eqnarray}
where $v^{\mu}$ is the four-velocity of the charmonium multiplet $J$. However, even in the boosted frame, the structure of the interaction term remains unchanged. Although the transition matrix element only depends on the energy of photon, the transition rate can vary due to the kinematics. When $\eta_c$ carries large transverse momentum, and we go into the longitudinally co-moving frame of the charmonium states, we have the relation 
\begin{equation}
|{\bf p}_{f\perp}| \approx |{\bf p}_{i\perp}| \gg m_{J/\psi}, m_{\eta_c} \gg p_{fx}, p_{ix} 
\label{eq16}
\end{equation}
along with the delta function which imposes the conservation of transverse momentum in (\ref{eq9}). Thus, we can approximately write 
\begin{eqnarray}
E_{fi} &\approx& (m_{J/\psi}^2-m_{\eta_c}^2)/(2p_{i\perp}) ,
\nonumber \\
(E_i-p_{ix})^2 &\approx& p^2_{i\perp}+m^2_i-2p_{ix}p_{i\perp} .
\label{eq17}
\end{eqnarray}
Using these approximations, the relativistic transition probability becomes:
\begin{eqnarray}
|a_{fi}^{\rm rel}|^2 &=& \frac{g'^2}{4E_iE_f} \left(\frac{Ze\gamma b}{4\pi}\right)^2
\left(\frac{2E_{fi}\, K_1\left(\frac{E_{fi}R}{\gamma}\right)}{\gamma^2R}\right)^2
\nonumber \\
&\approx & \frac{g'^2}{4p_{i\perp}^2}\, \frac{Z^2\alpha\gamma^2 b^2}{4\pi}\, 
\left(\frac{2E_{fi}\, K_1\left(\frac{E_{fi}}{\gamma}R\right)}{\gamma^2R}\right)^2.
\label{eq18}
\end{eqnarray}
Compared with the nonrelativistic result (\ref{eq14}), the relativistic result (\ref{eq18}) contains a factor $p^2_{i\perp}$ in the denominator, which causes a significant suppression when the transverse momentum is much greater than the charmonium mass.

\section{Numerical results and conclusion}

Now, let us find some numerical results for Au+Au collisions at the highest RHIC energy. By taking $Z=79, b=R=7~{\rm fm}, m_c=1.4~{\rm GeV}, v=1$ and $\gamma=100$, and using (\ref{eq5}), we obtain the transition probability $|a_{fi}^{\rm QM}|^2 = 2.37\times 10^{-4}$ in the quark model. To derive the numerical result for the EFT calculation, we need to determine the numerical value of $\rho$. By computing the decay width of $J/\psi\rightarrow \gamma\eta_c$ in the framework of NRHH$\chi$PT and comparing the result with that derived in the quark model \cite{Voloshin:2007dx}, we find $\rho=0.147~{\rm GeV}^{-1}$ and thus $|a_{fi}^{\rm NR}|^2 = 2.37\times 10^{-4}$. If the relativistic correction is taken into account, we find $\rho=0.137~{\rm GeV}^{-1}$ and $|a_{fi}^{\rm rel}|^2 = 2.06\times 10^{-4}$. For the transition with large energy of the photon, the probability amplitude can also be computed in EFT framework by utilizing potential non-relativistic QCD(pNRQCD)\cite{Brambilla:2005zw}, while the result at leading order would be tantamount to that in the quark model. We thus conclude that the leading-order result for the magnetically induced transition probability from $\eta_c$ to $J/\psi$ is about $2\times 10^{-4}$. Unfortunately, this contributi
 on to $J/\psi$-production is hardly observable experimentally. 

We now consider some possible sources of corrections to this result. At the next-to-leading order, the coupling constant $\rho$ can be affected by the contribution of intermediate hadronic loops, which is indicated in the recent study of charmonium decays \cite{Li:2007xr,Guo:2010ak}. However, when matching to the experimental data of $J/\psi\rightarrow \gamma\eta_c$ decay, the correction will lead to further suppression on $\rho$, which makes the transition probability even smaller. In pNRQCD, the correction from the higher derivatives will result in the suppression of transition matrix element as well\cite{Brambilla:2005zw}. According to (\ref{eq14}), the transition probability will be independent of $\gamma$ in the ultra-relativistic limit ($\gamma\to\infty$) since $K_1(x)\approx x^{-1}+\mathcal{O}(x)$ as $x\rightarrow 0$. Increasing the collision energy further (e.~g.\ to LHC energies) will thus not change the result much. On the other hand, the result (\ref{eq18}) for large transverse momentum will give a more substantial suppression. We thus conclude that our quantitative estimate represents a robust upper bound.

In summary, we analyzed the contribution to $J/\psi$ production in relativistic heavy ion collisions through the magnetic dipole transition of $\eta_c$ due to the interaction with the strong magnetic field generated in the collision. We evaluated the transition probability in the framework of the quark model and in the EFT approach. The EFT approach shows that the polarization of the produced $J/\psi$ will be parallel to the direction of the magnetic field. Both approaches lead to comparable results, in which the transition probability is about $2\times 10^{-4}$, which is probably too small to be observed experimentally. We also considered the case when the $\eta_c$ and $J/\psi$ carry a large transverse momentum and found that the result is further suppressed. 

This work was supported in part by a grant from the U.S.~Department of Energy (DE-FG02-05ER41367).
\section*{References}
%\bibliography{etacgammatoJpsi}
\providecommand{\newblock}{}

\end{document}